\documentclass[12pt]{article}

\pdfoutput=1

\usepackage{graphicx}
\usepackage{amssymb}

\usepackage{bbm}

\usepackage{color}

\newcommand{\rf}[1]{(\ref{#1})}
\newcommand{\beq}{\begin{equation}}
\newcommand{\eeq}{\end{equation}}
\newcommand{\bea}{\begin{eqnarray}}
\newcommand{\eea}{\end{eqnarray}}

\newcommand{\e}{\mbox{e}}

\newcommand{\lam}{\lambda}

\renewcommand{\a}{\alpha}

\newcommand{\del}{\delta}

\newcommand{\dg}{\dagger}

\newcommand{\ra}{\rangle}

\newcommand{\hH}{{\hat{H}}}

\newcommand{\vac}{|0\ra}

\newcommand{\vaccu}{\langle{\rm vac}|}
\newcommand{\cuum}{|{\rm vac}\rangle}

\newcommand{\Wtre}{W^{(3)}}

\begin{document}

\begin{center}
\vspace{44pt}
{ \Large \bf CDT and the Big Bang\footnote{Presented at the 
3rd Conference of the Polish Society on Relativity. To appear in the proceedings.}}

\vspace{30pt}

{\sl J. Ambj\o rn}$\,^{a,b}$ and   {\sl Y. Watabiki}$\,^{c}$

{\small

\vspace{24pt}

$^a$~The Niels Bohr Institute, Copenhagen University\\
Blegdamsvej 17, DK-2100 Copenhagen \O , Denmark.

\vspace{8pt}

$^b$~Institute for Mathematics, Astrophysics and Particle Physics (IMAPP)\\ 
Radboud University,
Heyendaalseweg 135, 6525 AJ Nijmegen, \\The Netherlands.

\vspace{8pt}

$^c$~Tokyo Institute of Technology,\\ 
Dept. of Physics, High Energy Theory Group,\\ 
2-12-1 Oh-okayama, Meguro-ku, Tokyo 152-8551, Japan

}

\end{center}

\vspace{36pt}

\begin{center}
{\bf Abstract}
\end{center}

\vspace{6pt}

\noindent 
 We describe a CDT-like model where breaking of W3 symmetry will
lead to the emergence of time and subsequently of space.
Surprisingly the simplest such models which lead to higher 
dimensional spacetimes are based on the four ``magical'' Jordan 
algebras of 3x3 Hermitian  matrices 
with  real, complex, quaternion and octonion entries, respectively. 
The simplest symmetry breaking leads to universes 
with spacetime dimensions 3, 4, 6, and 10.

\vfill

%\noindent
%--------------------------------------------------------------------------------

\newpage

\subsection*{Introduction}

String field theory is notoriously complicated, but so-called  
non-critical string field theory \cite{NCSFT,watabiki,aw} is a lot simpler.  
An even simpler version is the so-called CDT string field theory \cite{GSFT}.
The starting point is the continuum limit of two-dimensional 
causal dynamical triangulations (CDT) \cite{al}, a limit which is two-dimensional 
Ho\v{r}ava-Lifshitz quantum gravity if spacetime topology is trivial \cite{agsw}.
However, here we are interested in a generalized CDT where spacetime 
topology can change \cite{GCDT}. This theory describes the dynamics of 
topology changes of two-dimensional spacetime. One has  
creation and annihilation operators, $\Psi^\dg(L)$ and 
$\Psi(L)$, for spatial universes of length $L$ and the Hamiltonian involves 
terms like
\beq\label{zj1}
\Psi^\dg(L_1)\Psi^\dg(L_2)\Psi(L_1+L_2),\quad\quad
\Psi^\dg(L_1+L_2)\Psi(L_2)\Psi(L_1),
\eeq
which describe the splitting of a spatial universe in two, and the merging of two 
spatial universes into one such universe. 

Although string field theory deals with splitting and merging of spatial universes,
at the end of the day it tells us surprising little about the creation of 
a universe from ``nothing'', i.e. Big Bang. This led us to look for some 
general symmetry, the breaking of which could result in a Big Bang scenario. 
It is possible to derive the CDT string field Hamiltonian by starting out with a
$W^{(3)}$-symmetric theory which {\it a priori} has no spacetime interpretation.
However, when the  $W^{(3)}$ symmetry is broken ``time'' and correspondingly
the (CDT) Hamiltonian will ``emerge'', but in such a way that also space can be created
from ``nothing''. Below we will shortly describe how this is realized. However, in this way
one only obtains a one-dimensional space. We will then argue that this one-dimensional
scenario can be generalized to  2,3,5 and 9 spatial dimensions if one 
considers $W^{(3)}$ algebras with intrinsic symmetries related to one of the so-called
``magical'' Jordan algebras.

\subsection*{Why $\Wtre$ ?}

When one discusses splitting and joining of strings one encounters terms like \rf{zj1}.
In the case of ordinary non-critical string field theory this lead to a special $\Wtre$ algebra
which ensures that the partition function is a $\tau$-function \cite{kdv}.
In the case of CDT string field theory we now  promote the $\Wtre$-symmetry to the 
starting principle.  

The formal definition of  $W^{(3)}$ operators in terms of 
operators $\a_n$ satisfying   
\beq\label{jx10} 
[\a_m,\a_n] = m \,\del_{m+n 0},
\eeq
is the following
\beq\label{jx12}
W^{(3)}_n = \frac{1}{3} \sum_{k,l,m} :\!\a_k\a_l\a_m\!: \del_{k+l+m,n}.
\eeq 
%\beq\label{jx11}
%a(z) = \sum_{n \in \mathbbm{Z}} \frac{\a_n}{z^{n+1}},
%\qquad
%W^{(3)}(z) = \frac{1}{3}\! :\!\a(z)^3\!:
%\;= \sum_{n\in Z} \frac{W^{(3)}_n}{z^{n+3}}.
%\eeq
The normal ordering $:\!\!(\cdot)\!\!:$ 
refers to the $\a_n$ operators ($\a_n$ to the 
left of $\a_m$ for $n>m$. This ordering is opposite to the one  conventionally used.
See \cite{aw} for the motivation for such a choice).

We then define the ``absolute vacuum'' $|0\ra$ by the following condition:
\begin{equation}\label{jy2}
\a_n |0\rangle =  0,\quad n < 0,
\end{equation}
and the so-called $W$-Hamiltonian ${\hH}_{\rm W}$ by 
\beq\label{jy3}
{\hH}_{\rm W} := - W^{(3)}_{-2} =
-\,\frac{1}{3} \sum_{k, l, m}
:\!\alpha_k \alpha_l \alpha_m\!: \delta_{k+l+m,-2}.
\eeq
Note that ${\hH}_{\rm W}$ does not contain any coupling constants.

It was  shown in \cite{awBB} that by introducing a coherent state, 
which is an eigenstate of $\a_{-1}$ and $\a_{-3}$ and which 
we denoted the ``physical'' vacuum state $\cuum$, ${\hH}_{\rm W}$ was 
closely related to the CDT string field Hamiltonian   $\hH$. 
We thus defined 
\beq\label{zj5}
\cuum \propto \e^{ \lam_1 \a_{1} + \lam_3 \a_3} \vac,
\eeq
and we have 
\beq\label{zj6}
\a_{-1} \cuum =\lam_1 \cuum, \qquad \a_{-3} \cuum= 3 \lam_3 \cuum.
\eeq
The main point is the  following: because $\vaccu \a_n \cuum$
is different from zero for $n=-1$ and $n=-3$, 
${\hH}_{\rm W}$ will now contain terms only involving two operators $\a_l$. 
These terms can act like quadratic terms in $\hH$. 
At the same time the cubic terms left in ${\hH}_{\rm W}$ 
will act like the interaction terms in $\hH$, 
resulting in joining and splitting of universes. 
Finally, the expectation values of $\a_{-1}$ and $\a_{-3}$ 
determine the coupling constants of $\hH$.
More precisely one has \cite{awBB}
 \beq\label{zj2}
 {\hH}_{\rm W}  \propto \hH + c_4 \phi_4^\dg +c_2 \phi^\dg_2,
 %+ c_1 \phi_1^\dg + c_0,
 \eeq
where $\hH$ is the CDT string field Hamiltonian. 
$c_4$ and $c_2$ are constants.
%determined by $\lambda_1$ and $\lambda_3$ uniquely.
The creation operators $\phi^\dg_n$ are the $\a_n$, $n > 0$, 
while annihilation operators $\phi_n$ are 
related to $\a_n$, $n < 0$, except that 
$\phi_1$ and $\phi_3$ are shifted by eigenvalues given in eq.\ \rf{zj6}, 
such that $\phi_n \cuum =0$. 
$\hH$ is normal ordered such that $\hH \cuum = 0$.

By breaking the $\Wtre$ symmetry one can thus obtain CDT string field theory
except for one important point: {\it the vacuum is not stable}. 
The terms  $c_4 \phi_4^\dg +c_2 \phi^\dg_2$ %$+ c_1 \phi_1^\dg $ 
cause universes of infinitesimal length to be created 
and the non-interacting part of $\hH$, 
which explicitly can be written as 
\beq\label{zj3}
 \hH_{0} =- \sum_{l=1}^\infty \phi_{l+1}^\dagger l \phi_l
+ \mu \sum_{l=2}^\infty \phi_{l-1}^\dagger l \phi_l,
\eeq
might expand such an infinitesimal length  space to macroscopic size. The precise 
relation between the operators $\phi_l,~\phi^\dg_l$ and the operators 
$\Psi(L),~\Psi^\dg(L)$ which  annihilate and create spatial universes of 
macroscopic length $L$ is as follows 
\beq\label{jj1}
\Psi^\dagger(L) = \sum_{l=0}^\infty \frac{L^l}{l!} \; \phi^\dg_l,
\eeq
and thus operators $\phi^\dg_l$ with small $l$ create only infinitesimal size 
spatial universes.

\subsection*{Generalization to higher dimensions}

Above the spatial universe created from nothing was one-dimensional. 
We can introduce higher dimensional spaces by attaching intrinsic ``flavors''
to different spatial directions. In addition we want to be able to rotate these 
flavors into each other. This leads to so-called extended $\Wtre$-algebras, which
again are related to Jordan algebras \cite{romans}. Surprisingly it turns out that only 
the four so-called magical Jordan algebras allow us to make  symmetry 
breakings which lead to  CDT-like Hamiltonians of the kind considered above.

We find for the 
$\Wtre$ Hamiltonian the expression
\beq\label{jordan4}
{\hH}_{\rm W} := -W_{-2}^{(3)} =
-\,\frac{1}{3} \sum_{k, l, m} \sum_{a, b, c} d_{abc}
:\!\alpha_k^{(a)} \alpha_l^{(b)} \alpha_m^{(c)}\!: \delta_{k+l+m,-2},
\eeq
where $d_{abc}$ are the structure constants for the Jordan algebras.
The magical Jordan algebras are Hermitean $3\times 3$ matrices 
$H_3(\mathbbm{F})$, where $\mathbbm{F}$ denotes 
$\mathbbm{R}$, $\mathbbm{C}$, $\mathbbm{H}$ and $\mathbbm{O}$ (the real numbers,
the complex number, the quaternions and the octonions) and 
the structure constants are related to the standard Gell-Mann $d_{abc}$ 
for $H_3(\mathbbm{C})$ in a simple way. 
Again the model only allows a spacetime interpretation 
after choosing a specific coherent state. 
Here we discuss only the simplest, interesting choice,
namely breaking in the 8-direction (in the notation of Gell-Mann). 
Instead of \rf{zj5} and \rf{zj6} we can choose 
\beq\label{zj5b}
\cuum_{8} \propto
\e^{ \lam_1^{(8)} \a_{1}^{(8)} + \lam_3^{(8)} \a_3^{(8)}}\vac,
\eeq
and we have 
\beq\label{yj6}
\a_{-1}^{(8)}  \cuum_{8} =\lam_1^{(8)}  \cuum_{8},
\qquad
\a_{-3}^{(8)}  \cuum_{8} = 3 \lam_3^{(8)}  \cuum_{8}.
\eeq

Again one obtains an unstable vacuum. The kinetic, non-interacting part
of the Hamiltonian (the part corresponding to \rf{zj3}) has 
coefficients $d_{ab8}$ and in fact only 
 coefficients $d_{aa8}$ are different from zero. 
One can argue \cite{aw3} that in order for the non-interacting part of the Hamiltonian to
allow a universe to expand from infinitesimal size to macropscopic size  
one has to demand that the coefficients $d_{aa8} > 0$.
When we then look at the four magical algebras, we have 
for $H_3(\mathbbm{R})$ two indices $a$ where $d_{aa8} = 1/\sqrt{3}$. 
For $H_3(\mathbbm{C})$ we have three indices $a$ where $d_{aa8} = 1/\sqrt{3}$.
For $H_3(\mathbbm{H})$ we have five indices $a$ where $d_{aa8} = 1/\sqrt{3}$ 
and finally for $H_3(\mathbbm{O})$ 
we have nine indices $a$ where $d_{aa8} = 1/\sqrt{3}$. The rest of the $d_{aa8}$ are 
non-positive.
The symmetry breaking 
corresponds to breaking the automorphism group of the $\Wtre$ algebras 
from $SO(3)$ to $SO(2)$, 
from $SU(3)$ to $SO(3)$, 
from {\it U}$Sp(6)$ to $SO(5)$ and
finally from $F_4$ to $SO(9)$. 
The extended spacetime dimensions will be (including the time) 
2+1, 3+1, 5+1 and 9+1 
which are the dimensions of the classical superstrings.
It is an interesting question if one can  consider the flavor indices $a$  of space
as reflecting matter fields partly integrated out. It opens for the exciting possibility that 
the matter content is unique.

\subsection*{Acknowledgments}

JA and YW acknowledge support from the ERC-Advance grant 291092,
``Exploring the Quantum Universe'' (EQU).

\end{document}